\documentclass[aps,prl,floats,twocolumn]{revtex4}
 \usepackage{graphicx}
\usepackage{amssymb}
\begin{document}
\title{Recombination dramatically speeds up evolution of finite populations}

\author{Elisheva Cohen} 
\affiliation{Dept. of Physics, Bar-Ilan  University, Ramat-Gan, IL52900 Israel}
\author{David A. Kessler}
\affiliation{Dept. of Physics, Bar-Ilan University, Ramat-Gan, IL52900 Israel}
\author{Herbert Levine} 
\affiliation{Center for Theoretical Biological
  Physics, University of California, San Diego, 9500 Gilman Drive, La
  Jolla, CA 92093-0319} 
\date{\today}
\begin{abstract}
We study the role of recombination, as practiced by genetically-competent bacteria, in speeding up Darwinian evolution. This is done by adding a new process to a previously-studied Markov model of evolution on a smooth fitness landscape; this new process allows alleles to be exchanged with those in the surrounding medium. Our results, both numerical and analytic, indicate that for a wide range of intermediate population sizes, recombination dramatically speeds up the evolutionary advance.
\end{abstract}
\maketitle 

Recombination of genetic information is a common
evolutionary strategy, both in natural systems~\cite{review} as well
as in-vitro molecular breeding~\cite{stemmer}.  This idea is also employed
in genetic programming~\cite{baum}, a branch of computer
science which aims to evolve efficient algorithms.  Given all
this, it is surprising that we still do not have a good understanding
of the conditions under which the benefits of recombination outweigh
the inevitable costs.

Of course, there is a large literature on recombination, dating back
to the ideas of Muller~\cite{muller} and Knodrashov~\cite{kondrashov}.
One line of recent work
focuses on two loci genomes and considers whether or not
recombination would be favored; possible mechanisms include (weak)
negative epistasis (the fact that the reproduction rate is not just
the sum of the individual rates) or negative linkage disequilibrium
(the lack of independence of the allele distribution in a finite
population) or some combination thereof~\cite{barton}. Others look at
how the (static) genetic background in which a mutation arises will
affect fixation probabilities (``clonal interference"), comparing
these with or without recombination~\cite{charlesworth}. In both of
these methods, only one or two mutations at a time are ``dynamic", a
situation unlikely to be true for rapidly evolving microorganisms.  In
contrast, our analysis considers a large number of contributing loci.
 
In this paper, we study recombination in the context of a simple
fitness landscape model~\cite{evol-prl,evol-papers,rouzine} which has proven
useful in the analysis of laboratory scale evolution of viruses and
bacteria~\cite{lab-evol}.  The specific type of recombination we consider is based
on the phenomenon of bacterial competence~\cite{competence}.  Here,
bacteria can import snippets of DNA from the surrounding medium;
presumably these are then homologously recombined so as to replace the
corresponding segment in the genome.  This behavior is 
controlled by of a cellular signaling system that ensures that recombination
only occurs under stress.  The details of the DNA importation and the
aforementioned control has convinced most
biologists~\cite{levin,redfield} that competence is an important
survival strategy for many bacterial species.
        
Our model consists of a population of $N$ individuals each of which has a
 genome of $L$ binary genes.  An individual fitness depends additively on 
the genome $
x \ = \ \sum_{i=1}^L S_i$ with $S=0,1$.
Evolution is implemented as a continuous time Markov process in
which individuals give birth at rate $x$ and die at random so as to
maintain the fixed population size.  Every birth allows for the
daughter individual to mutate each of its alleles with probability
$\mu _0$ giving an overall genomic probability of $\mu = \mu _0 L$.

The last part of our Markov process concerns the aforementioned
recombination.  At rate $f_s L$, an individual has one of its genes
deleted and instead substitutes in a new allele from the surrounding
medium; the probability of getting a specific $S$ is just its
proportional representation in the population.  This mimics the
competence mechanism as long as the distribution of recently deceased
(and lysed) cells is close to that of the current population; this
should be the case whenever the random killing due to a finite
carrying capacity is the most common reason for death.
\begin{figure}
  \includegraphics[width=.4\textwidth]{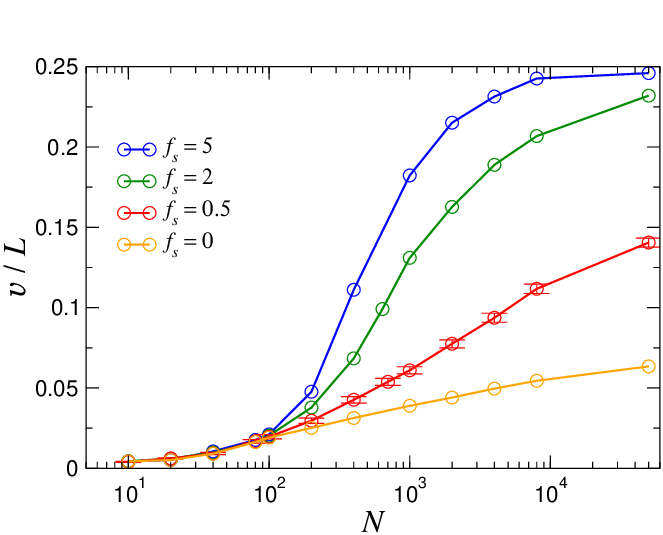}
  \includegraphics[width=.4\textwidth]{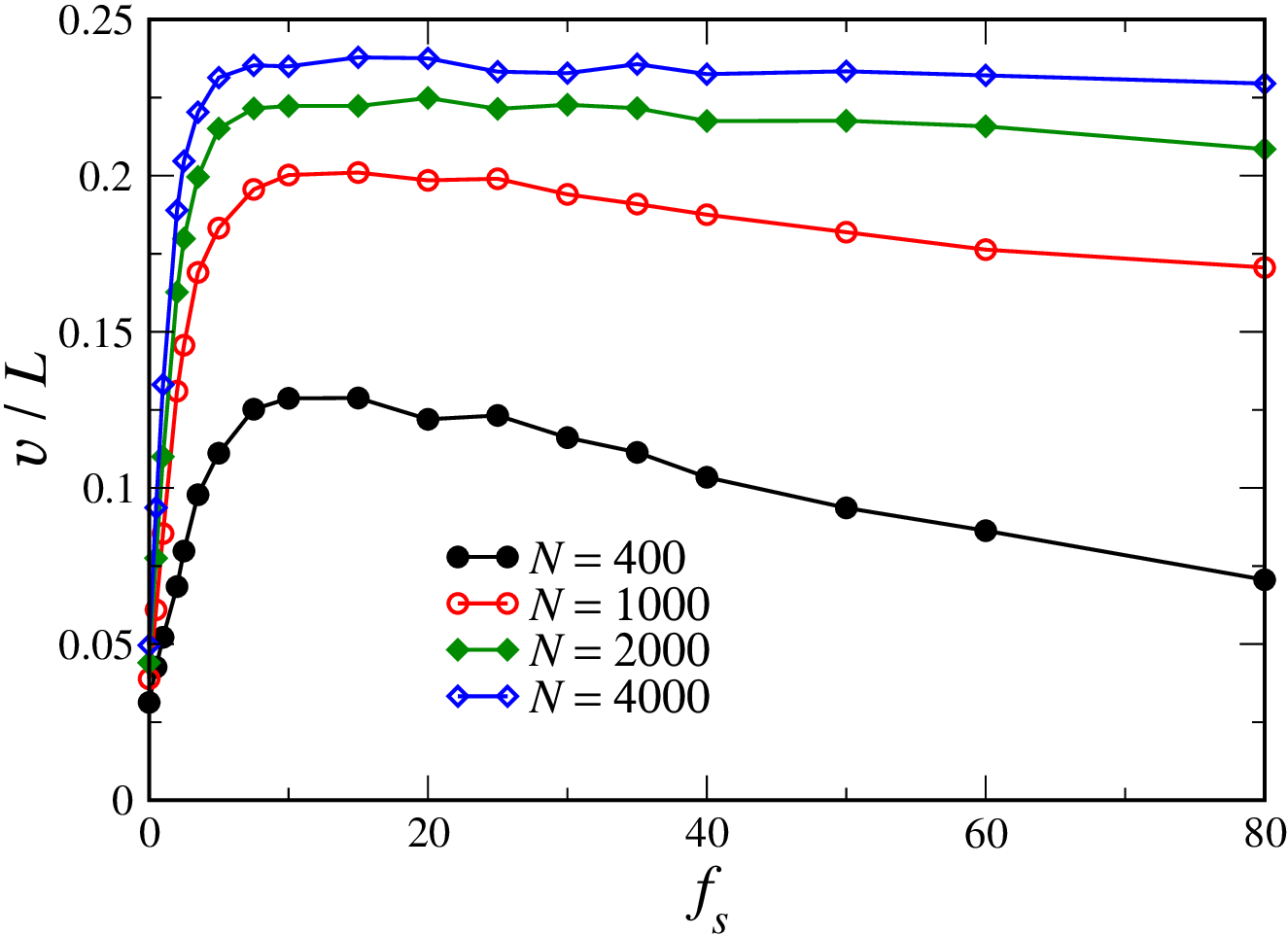}
\caption{Velocity (averaged over 200 samples) measured between $x=95$ and $x=105$ starting with an average fitness of $50$.  $L=200$, $\mu=0.1$. a) $v$ 
measured as a function of $N$, for various $f_s$. Error bars are shown for
one value of $f_s$, and are typical of all the data.  
b) $v$ measured as a function
of $f_s$ for various $N$. As $N$ increases, the velocity saturates at an $f_s$-independent value}
\label{fig1b}
\end{figure}   
In Fig. 1a, we show simulation results for the ``velocity'', i.e., the 
rate of fitness
increase, (at one representative point on the landscape) for different
O(1) values of $f_s$ (the recombination probability per time per
site), as a function of $N$.  At very small population sizes,
recombination has little effect, since there is no population
diversity upon which to act.  Each of the curves rises sigmoidally
to a saturation value at very large $N$ which is again roughly independent
of the recombination rate (see Fig 1b and later). Because the population scale for this rise
is a strongly decreasing function of $f_s$,
recombination at intermediate $N$ can give a dramatic speedup of the
evolution.  It is worth mentioning that this basic result is
qualitatively consistent with recent experiments~\cite{nature} in
microorganism evolution which demonstrate an increase in the efficacy
of recombination as the population size is increased (starting from
small); we should note however that the details of recombination in
the experimental systems are different than those underlying
bacterial competence.

Can one understand these simulation results?  At small $N$, we can
appeal to previous results for this model~\cite{evol-papers} that show that the
population variance scales as $\mu N$. One would therefore expect the
small $N$ breakpoint where the curves diverge to be roughly at $N=
1/\mu$; this is consistent with the data in Fig. 1a and we have
checked this simple scaling with mutation rate (data not shown).
Another small $N$ effect becomes evident if the simulations are
extended to much larger $f_s$ values, as shown in Fig. 1b.  Now, the
velocity begins a slow decline at too large $f_s$, due to the
recombination causing a loss of diversity as various
sites get locked into specific alleles.

The behavior at larger $N$, past the inflection point of the
velocity curves in Fig 1a and in the rising segments of the curves in
Fig 1b, is much less trivial. To make progress, we start by
assuming that the subpopulation at some particular
fitness $x$ has equal distributions at each site of the genome. This
assumption means, of course, that selecting at random an allele at any
site gives a chance $x/L$ of getting $S=1$ and $1-x/L$ of getting
$S=0$.  Then, one can write down an equation for the infinite
population size limit which directly determines the fitness
distribution function,
\begin{eqnarray}
\frac{d P_x (t)}{dt}  =  (x-\bar{x}) P_x (t) 
 +  \mu  \left[ \frac{(x+1)^2}{L} P_{x+1} (t) +  \right.  & \nonumber \\ 
 \left.(x-1) \left( 1-\frac{x-1}{L} \right)  P_{x-1}(t)  -xP_x (t) \right]  &
 \nonumber \\
 -  f_s \left[ \left( 1- \frac{\bar{x}}{L} \right) \frac{x}{L} P_x (t)
 + \frac{\bar{x}}{L} \left( 1-\frac{x}{L} \right) P_x (t) \right. 
 -  & \nonumber \\
  \left.  \left( 1- \frac{\bar{x}}{L} \right) \frac{x+1}{L} P_{x+1} (t)
 + \frac{\bar{x}}{L} \left( 1-\frac{x-1}{L} \right) P_{x-1} (t) \right] &
 \label{MFE}
\end{eqnarray}
The first two terms are standard and reflect the birth-death process
and the genomic mutation respectively; the explicit form of the
mutation term arises from considering the probability of an individual
with fitness $x$ giving birth (rate $\sim x$), mutating ($\sim \mu$),
and hence going either up ($\sim (1-x/L)$, the number of currently
bad alleles) or down ($\sim x/L$, the number of good alleles).
The last term is new and reflects the role of recombination.  With the
aforementioned assumption, the probability that an individual of
fitness $x$ will have its fitness altered is proportional to the recombination
rate, $f_s$ times the probability of either: a) deleting a bad
allele ($1-x/L$) and picking up a good one ($\bar{x}/L$); or b)
deleting a good allete ($x/L$) and picking up a bad one
($1-\bar{x}/L$).

Before using this equation (and its modification for finite
$N$ effects; see below) to analyze the numerical results, we need to
test the underlying equi-distribution assumption.  To do this,
we generated a population of $N=1000$ at $f_s=2$, and let it evolve
until reaching $\bar{x}=75$, for $L=100$.  We then measured the
respective probabilities for a recombination event to increase or
decrease the fitness, based on the fitness $x$ of the chosen
individual.  As shown in Fig. 2, our theoretical expression has
the correct functional dependence, although it
overestimates these actual probabilities by roughly a fixed amount.  This
overestimate is due to the fact that individual sites have less
diversity than is predicted, a remnant of the aforementioned loss-of-diversity effect.
Notwithstanding the error (which we find
decreases as $N$ increases), this comparison gives us confidence that
the above equation can account semi-quantitatively for the
recombination process.
\begin{figure}
  \includegraphics[width=.4\textwidth]{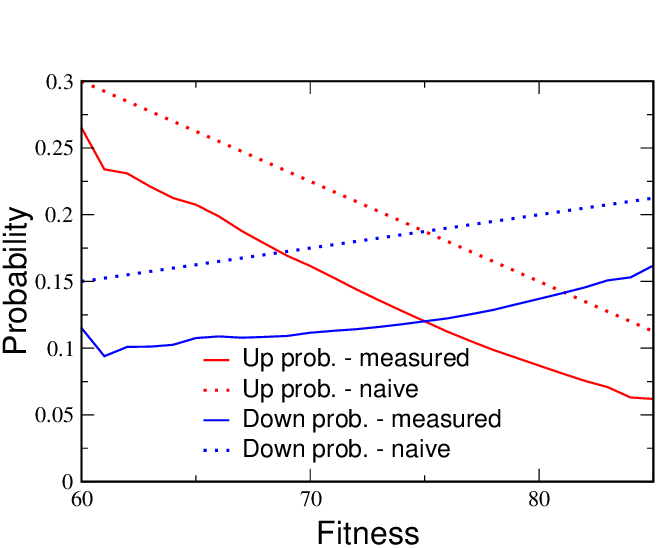}
\caption{The probability of making an up and down move due to recombination
as a function of individual fitness, when the average fitness of the population
was 75.  For this run, $N=1000$, $L=100$, $f_s=1$, $\mu=0.1$. The ``naive''
probabilities are those derived assuming equal distribution of alleles at
every locus.}
\label{fig2}
\end{figure}

\begin{figure}
\includegraphics[width=.4\textwidth]{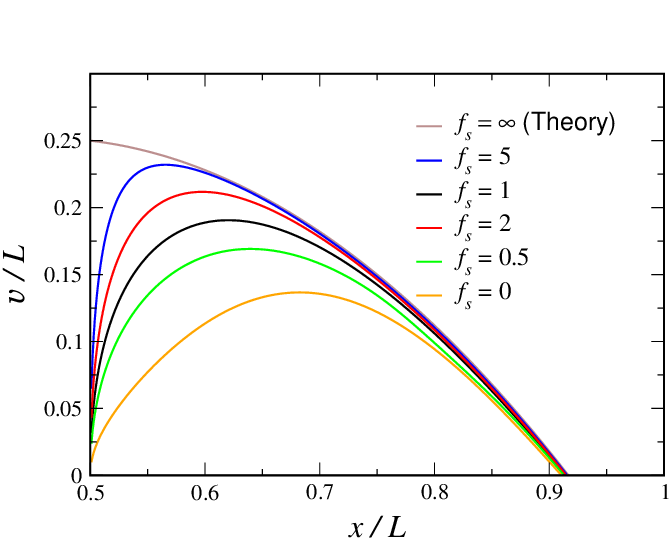}
\caption{Velocity vs. average fitness for simulations of the non-cutoff
MFE (Eq. (\protect{\ref{MFE}})) for various $f_s$.  
Parameters are $\mu=0.1$, $L=1000$, initial fitness $x_0=500$. The curve
for $f_s=\infty$ is taken from Eq. (\protect{\ref{inf-fs}}).}
\label{fig3a}
\end{figure}

In Fig. 3a, we show the results of solving Eq. (1) numerically for a
variety of $f_s$ values. At non-zero $f_s$,
the fitness rapidly approaches a universal trajectory which is $f_s$
independent; only the rate of approach varies.  Hence, the amount of
recombination is of minor importance if $N$ is large enough for this
mean-field theory to apply. We can explain this 
by noting that the recombination term
on its own tends to make the population relax to a distribution that
satisfies the equation
\begin{eqnarray}
0 &=& ( 1- \frac{\bar{x}}{L} ) \frac{x+1}{L} P_{x+1} (t) 
+ \frac{\bar{x}}{L} ( 1-\frac{x-1}{L} ) P_{x-1} (t)  \nonumber \\
  &\ & \  -   \  ( 1- \frac{\bar{x}}{L} ) \frac{x}{L} P_x (t)
 + \frac{\bar{x}}{L} ( 1-\frac{x}{L} ) P_x (t)    
   \label{binomial} 
\end{eqnarray} 
It is easy to verify that choosing $P$ to be binomial, $B(L,\bar{x}/L)$, 
satisfies this requirement. The evolutionary dynamics can then be determined
 by multiplying both sides of the MFE, Eq. (\ref{MFE}), by $x$ and then summing over  $x$,
 thereby computing the time derivative of $\bar{x} 
\equiv pL$. This yields
\begin{equation}
\dot{p} = p(1-p) +\mu ( p- \frac{2}{L} (p(1-p)+Lp^2)) 
\label{inf-fs}
\end{equation}
 Solving, we obtain for the case of initial $p(0)=1/2$
$$
p(t) \ = \ \frac{1+\mu (1-2/L)}{1+2\mu -2\frac{\mu}{L}
 +(1-2\frac{\mu}{L}) \exp {((-1+\mu -2\frac{\mu}{L })t)}}
$$
The final state is reached in an O(1) time and this indeed is
quite rapid evolution; this analytic curve is included in Fig. 3. Now, the fact that
recombination attempts to enforce a binomial distribution but
otherwise does not directly change the rate of evolutionary advance
explains why it has little consequence in the $N \rightarrow \infty$
mean-field limit.  Essentially, the pure mutation-selection problem
will, up to small corrections if $L$ is large, also give rise to a
binomial distribution which therefore self-consistently solves the
entire equation. To see this, we replace the birth rate factors in the
mutational part of the MFE by the constant rate $\bar{x}$; this
introduces an error of $O(\frac{x-\bar{x}}{L})$, which becomes
$O(L^{-1/2})$ were we to have a binomial distribution. Then, we can
directly check that the same binomial anstaz solves the $f_s=0$
time-dependent MFE, giving rise to
$\dot{p} = p(1-p) +\mu p ( 1-2p)$;
this agrees with the above equation for large $L$. Hence, the only role for
 recombination is to cause the system to dynamically {\em select} this
 particular solution of the mean-field theory; the value of
 $f_s$ makes no difference, once we are past the transient period. 
 
 We have now explained why the large $N$ saturation value in Fig. 1a
 is roughly $f_s$ independent. The remaining issue concerns the
 critical value of $f_s ^* (N)$ at which the system reaches the
 plateau (see Fig 1b); the previous argument suggests that as $N
 \rightarrow \infty$, $f_s^* \rightarrow 0^+$. This value is of
 crucial importance, as it represents the amount of recombination
 needed for a {\em finite} population to achieve the maximal rate of
 evolution. Studying this requires inclusion of finite population
 effects in the evolution equation, for which we employ a heuristic
 cutoff approach which has been shown to be accurate in a variety of
 previous investigations~\cite{cutoff,prl}. In detail, we replace the first part of the
 mean-field equation (MFE) with the alternate form
\begin{equation}  
\frac{d P_x (t)}{dt}  =  (x \theta (P_x-P_c) -\lambda) P_x (t)
 \label{cutoff-MFE}
\end{equation} where $\lambda$ is chosen to satisfy population conservation
$$
\lambda =  \int dx \, xP_x \theta(P_x-P_c)
$$
and $P_c$ is a cutoff of order $1/N$.
Fig. 4a compares the time evolution of the stochastic system with that
 predicted by the cutoff MFE, showing reasonable agreement.
Finally, Fig 4b shows the desired effect, namely the fact that the transition
 point to rapid evolution is a decreasing function of $\ln N$.
\begin{figure}
  \includegraphics[width=.23\textwidth]{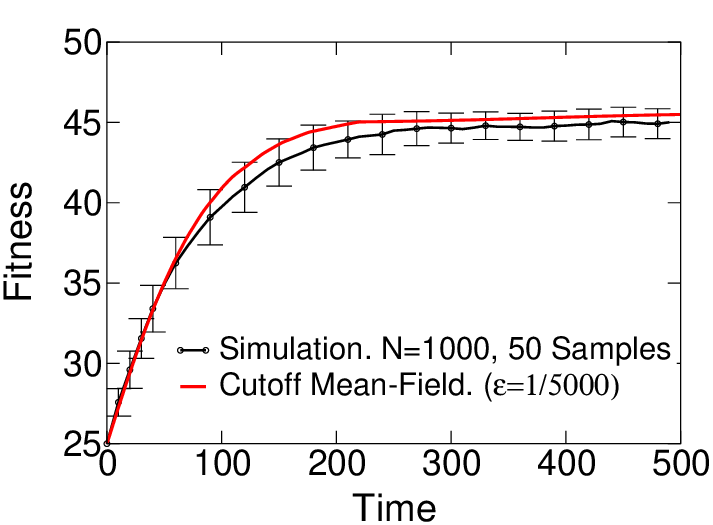}
\includegraphics[width=.23\textwidth]{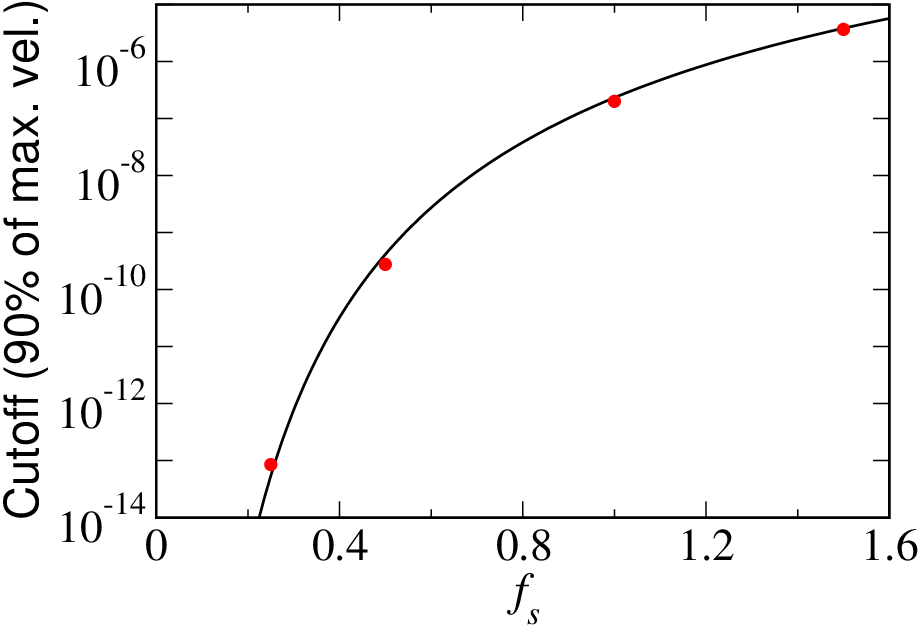}
\caption{a)Fitness as a function of time from an average over 50 runs, 
$N=1000$,
compared to a numerical solution of the cutoff MFE with cutoff $P_c=1/5000$. 
$\mu=0.1$, $f_s=2$, $L=50$, $x_0=25$. b) Cutoff for which $v$ is 90\%
of its maximal (cutoff=0) value as a function of $f_s$.  $L=200$, $\mu=0.1$,
Velocity measured between $x=95$ and $x=105$, with initial $x=50$.  Solid curve
is the theoretical prediction, Eq. \protect{\ref{scaling}}.}
\label{fig4a}
\end{figure}

Why does finite $N$ matter in this manner? It is easy to check that
the cutoff term has no consequential effect as long as the
distribution remains binomial. The real breakdown in the previous
analysis occurs when $N$ becomes small enough that the variance (and
hence the rate of fitness advance) saturates at $\ln {N}$ instead of
$L$. This transition means that the mutation-selection balance is not
consistent with the binomial. The simplest way to make an estimate of
the critical $N$ is to compare the calculated rate of mean fitness
advance $L\dot{p}$ based on the binomial distribution, with that to be
expected when finite $N$ effects are dominant. To estimate the latter,
we notice that the recombination term can be thought of as containing
both a drift piece and a diffusion piece
$$
+ \left[ V P \right] ' + \left[ DP \right] '' 
$$
where $'$ refers to the finite difference operator and $V_x =
\frac{x-\bar{x}}{2L}$, $D_x = \frac{x+\bar{x}}{L} - \frac{2 x
  \bar{x}}{L^2}$. The drift term is small, because $x-\bar{x}$ is a
power of $\ln N$ which is assumed much less than $L$; hence, the most
important effect is that of increased diffusion. This in fact
appears to be the secret behind the efficacy of recombination in this
model, namely that it acts to increase variation just like an
increased mutation rate but without a mean drift term, aka the
"mutational load". The diffusion coefficient is finite as long as we
are not near $\bar{x} =L$. Assuming recombination dominates, we can
use the results of previous analyses of the mutation-selection problem
with $f_s L$ substituted for the genomic mutation rate $\mu \bar{x}$.
From ref. \cite{evol-prl} , the velocity under this assumption scales
as
$$
v \sim (f_s L) ^{2/3} \ln ^{1/3} N
$$
Equating this to the previous velocity result, the predicted critical value of
 $f_s$ at which the system crosses over to rapid evolution is predicted to
 scale as
\begin{equation}
f_s ^* \ \sim  \ \frac{L^{1/2}}{\ln ^2 N} \label{scaling}
\end{equation}
This is consistent with the data shown in the figure and indeed with the
 limited direct simulation data in Fig. 1b.

At this stage of our understanding, it is impossible to make any {\em
  quantitative} contact with experimental data. Nonetheless,
conceptual insights that emerge from our study seem to offer solutions
for some of the mysteries underlying bacterial competence. Our results
show that in the population range of interest for many microorganism
colonies, there is a huge potential benefit to be gained from
recombination; nevertheless too much recombination can hurt, as the specific
genes are too rapidly driven to the most common allele even if it is
not the beneficial one. This perhaps explains why recombination is so
heavily regulated via intercellular signaling. The mechanism behind
this benefit seems to be the increased rate of effective diffusion on
the landscape, similar to what would happen with an increased mutation
rate except that there is no significant extra load. Finally, we have
already mentioned that our results are consistent with recent
experiments; these could be extended to check the basic prediction of our approach regarding
the scaling of the needed rate versus population size (eq.
\ref{scaling}). Even more exciting would be the determination that the
signaling system is used for imposition of this result, measuring the
effective population by quorum sensing and feeding the information
into the competence pathway.

The work of HL has been supported in part by the NSF-sponsored Center
for Theoretical Biological Physics (grant numbers PHY-0216576
and PHY-0225630), and that of DAK and EC by the Israel Science Foundation.

\end{document}